\documentclass[%
superscriptaddress,
 amsmath,amssymb,
 aps,twocolumn,
floatfix,
]{revtex4-2}

\usepackage{graphicx}
\usepackage{fancyhdr}
\usepackage{dcolumn}
\usepackage{bm}
\usepackage{physics}
\usepackage[version=4]{mhchem}
\usepackage[colorlinks=True,linkcolor=red,citecolor=blue,urlcolor=blue]{hyperref}

\newcommand{\pba}{\ce{Pb10(PO4)6O}}
\newcommand{\pbcu}{\ce{Pb9Cu(PO4)6O}}
\newcommand{\pbx}{\ce{Pb$_{10-x}$Cu$_x$(PO4)6O}}

\newcommand{\mev}{\mathrm{\:meV}}
\newcommand{\kelvin}{\mathrm{\:K}}

\begin{document}


\title{Magnetic fluctuations in \ce{Pb9Cu(PO4)6O}}

\author{Makoto Shimizu}
\affiliation{Department of Physics, Graduate School of Science, Kyoto University, Kyoto 606-8502, Japan}

\author{Junya Otsuki}
\affiliation{Research Institute for Interdisciplinary Science, Okayama University, Okayama 700-8530, Japan}

\author{Harald O. Jeschke}
\affiliation{Research Institute for Interdisciplinary Science, Okayama University, Okayama 700-8530, Japan}

\date{\today}

\begin{abstract}
The hope that copper doped lead apatite {\pbcu} is a room-temperature superconductor has largely been dashed by global research efforts. Nevertheless, according to the current state of knowledge, the material has interesting magnetic properties, and research groups around the world have prepared high quality samples. We use a fluctuation exchange approximation (FLEX) approach to study the magnetic tendencies in {\pbcu}. We find ferromagnetic fluctuations very close to the filling of the stoichiometric compound which can be understood from Fermi surface nesting at the $M$ point.  This is similar to the one-band triangular lattice Hamiltonian at three-quarter filling. Interestingly, the special $k_z$ dependence of the {\pbcu} band structure makes it very sensitive to doping. Only slight charge doping switches between antiferromagnetic and ferromagnetic fluctuations. If the material could become superconducting, it might be easily switchable between singlet and triplet superconductivity.
\end{abstract}

\maketitle

\section{Introduction}

Copper doped lead apatite {\pbcu} has recently been put forward as a candidate for room temperature superconductivity at ambient pressure~\cite{Lee2023a,Lee2023b}, causing substantial excitement in the scientific community. Efforts at independent synthesis of the material and at reproduction of the experimental observations have been undertaken by many groups~\cite{Kumar2023,Hou2023,Guo2023}. Every failure to reproduce superconductivity~\cite{Kumar2023a} has added to the doubts concerning the interpretation of the original experiments. However, the suggestion that a well known phase transition in \ce{Cu2S}~\cite{Jain2023} which is a known impurity~\cite{Lee2023a} is responsible for the sudden change in resistivity around 104\,Kelvin in the mixed-phase {\pbcu} samples is rather convincing~\cite{Zhu2023}. Nevertheless, it is still an interesting question what the properties of the impurity free {\pbcu} material are. In particular, the half levitation~\cite{Lee2023b} that has been reproduced by some groups~\cite{Guo2023,Wu2023} points to interesting magnetic properties. The fact that excellent single crystals of {\pbcu} have been grown~\cite{Puphal2023} means that this question is now accessible to precise experiments.

The experimental claims for {\pbcu} have been met with a concerted response of the theoretical materials science community. Very fast density functional theory studies~\cite{Lai2023,Griffin2023,Si2023,Kurleto2023,Cabezas2023} have discussed the importance of flat bands and the shape of the Fermi surface. Several different tight binding models have been proposed~\cite{Si2023,Tavakol2023,Lee2023c} and analyzed with respect to topological properties~\cite{Hirschmann2023}. Three DFT+DMFT studies show that at the integer filling of the stoichiometric {\pbcu} compound, correlations open a gap~\cite{Korotin2023,Si2023a,Yue2023} in agreement with the transparent single crystals~\cite{Puphal2023}. Superconductivity has been theoretically studied using the two-dimensional $t$-$J$ model~\cite{Oh2023} and spin fluctuation theory~\cite{Witt2023}. First statements about the magnetism of {\pbcu} have been obtained~\cite{Sun2023}.

In this study, we will study the magnetic fluctuations in Cu doped {\pba} using the fluctuation exchange approximation (FLEX)~\cite{Bickers1989, Ikeda2010}. The stoichiometric {\pbcu} compound is described by a three-quarter filled two-orbital model. Its Fermi surface has been described as rugby ball shaped. We will take the three-dimensional nature of the compound seriously and work out similarity and differences between the electronic structure of {\pbcu} and a one-band model at different fillings on the triangular lattice. 

\begin{figure*}[hbt]
    \includegraphics[width=0.9\textwidth]{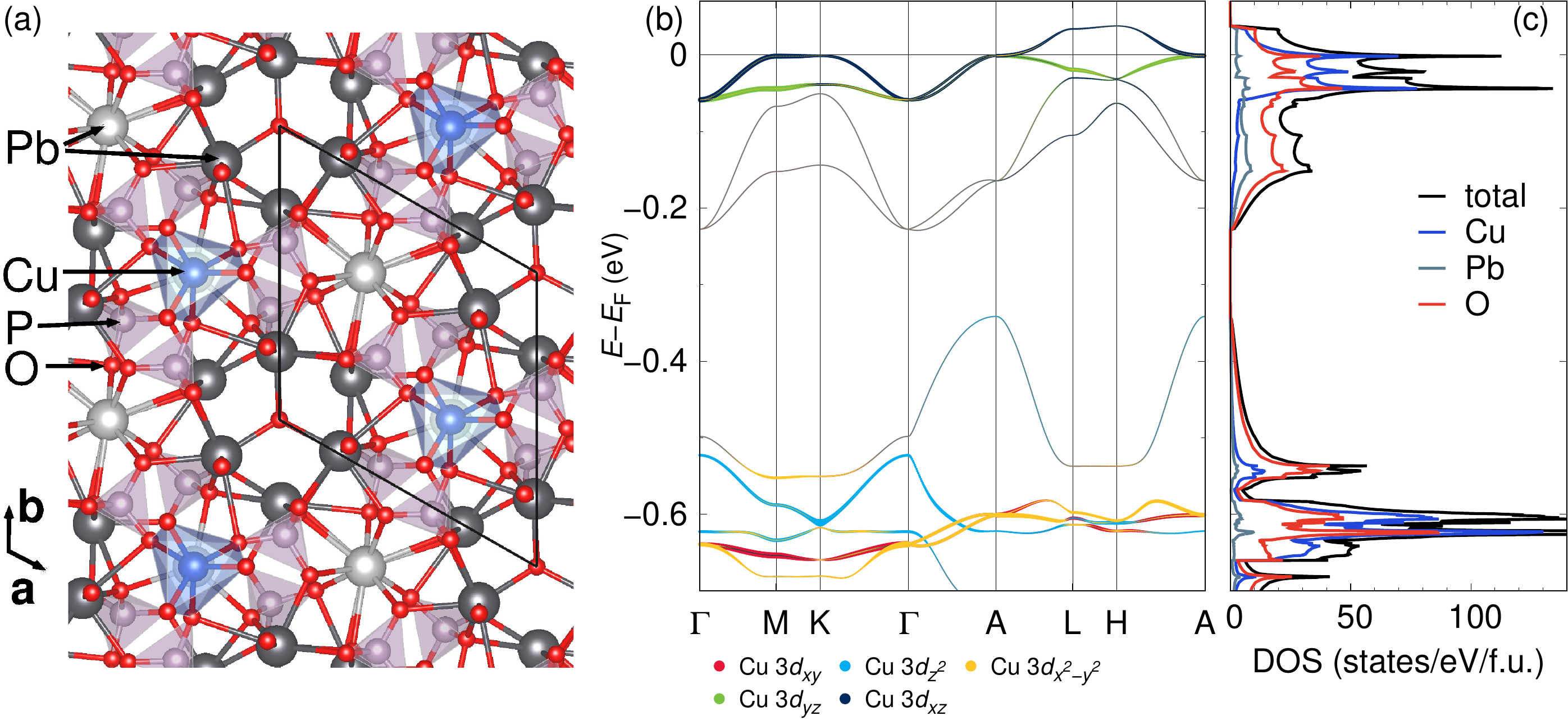}
    \caption{(a) Predicted simplified crystal structure of {\pbcu}. (b) Electronic band structure near the Fermi level at the GGA level with Cu $3d$ orbital character highlighted.}
    \label{fig:bands}
\end{figure*}

\begin{figure*}[hbt]
    \includegraphics[width=0.9\textwidth]{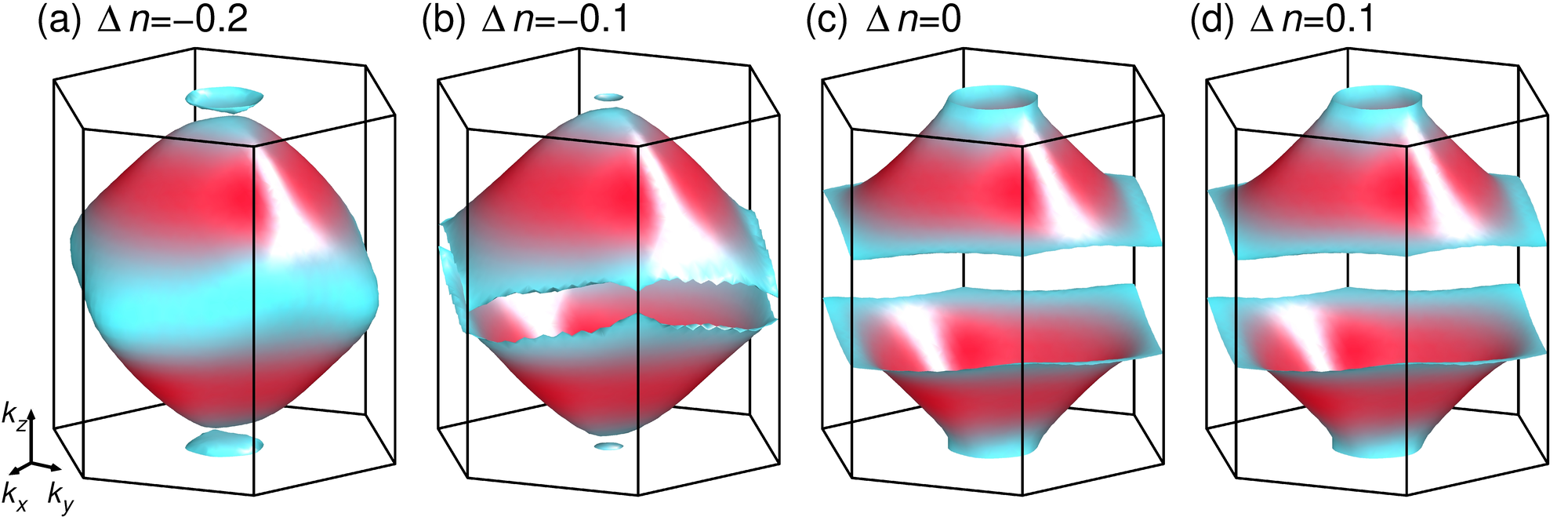}
    \caption{GGA Fermi surfaces of {\pbcu} for several doping levels $\Delta n = n - 7$. (c) is the Fermi surface of the stoichiometric compound, (a) and (b) correspond to hole doping of 0.2 and 0.1 electrons per formula unit, and (d) shows electron doping of 0.1 electrons. The color red (blue) indicates high (low) Fermi velocity.}
    \label{fig:3dfs}
\end{figure*}

\begin{figure*}[hbt]
    \includegraphics[width=\textwidth]{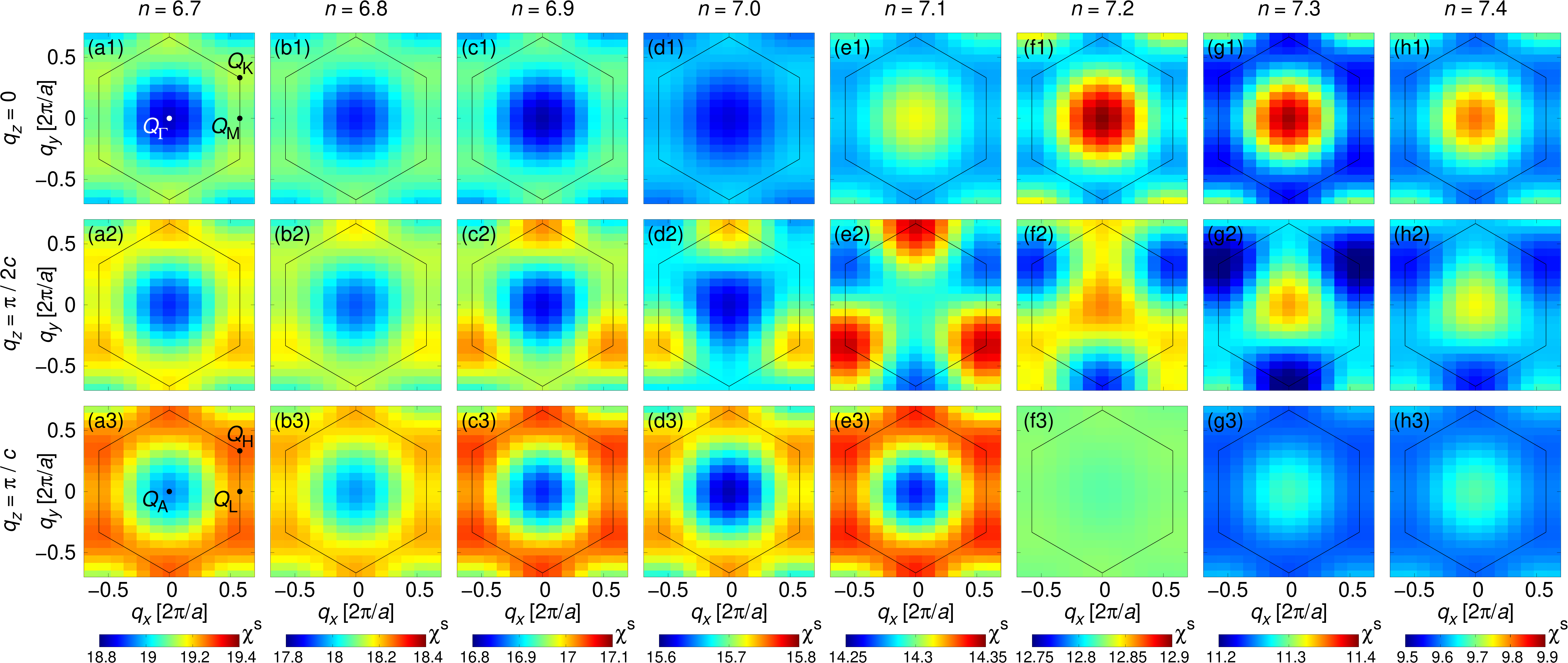}
    \caption{
      Cuts trough the static spin susceptibility $\chi^\mathrm{s}(\bm{q})$ for (a) $q_z = 0$, (b) $q_z = \pi/2c$ and (c) $q_z = \pi/c$ at $T=290\,\kelvin$ calculated within FLEX. Interaction parameters are $U=200\mev$, $V = U/2$ and $J = J' = U/4$.
      High-symmetry points are
      $\bm{Q}_{\Gamma} \equiv (0, 0, 0)$,
      $\bm{Q}_\mathrm{K} \equiv (1/\sqrt{3}, 1/3, 0)$,
      $\bm{Q}_\mathrm{M} \equiv (1/\sqrt{3}, 0, 0)$,
      $\bm{Q}_\mathrm{A} \equiv (0, 0, 1/2)$,
      $\bm{Q}_\mathrm{H} \equiv (1/\sqrt{3}, 1/3, 1/2)$,
      $\bm{Q}_\mathrm{L} \equiv (1/\sqrt{3}, 0, 1/2)$,
      in units of $2\pi/a$ for the $x$ and $y$ components and of $2\pi/c$ for the $z$ component. The high symmetry points are shown in (a1) and (a3).
    }
    \label{fig:chis_n6p7-7p4}
\end{figure*}

\begin{figure*}[hbt]
    \includegraphics[width=\textwidth]{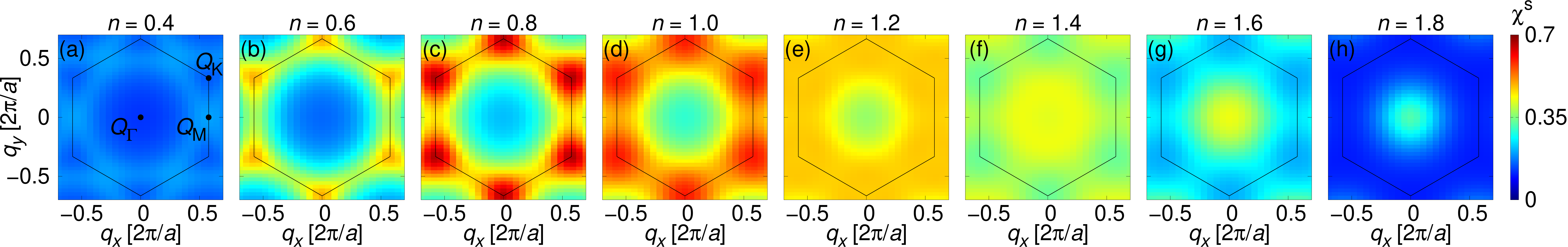}
    \caption{
      The static spin susceptibility $\chi^\mathrm{s}(\bm{q})$ of the triangular lattice at $T = |t| / 4$ calculated within FLEX. The interaction is $U = 9|t| = W$ where $t$ is the nearest neighbor hopping parameter and $W$ is the band width. We use a $32 \times 32 $ $\bm{k}$-mesh.
    }
    \label{fig:tri_chis_n0p4-1p8}
\end{figure*}

\begin{figure}[hbt]
    \includegraphics[width=\columnwidth]{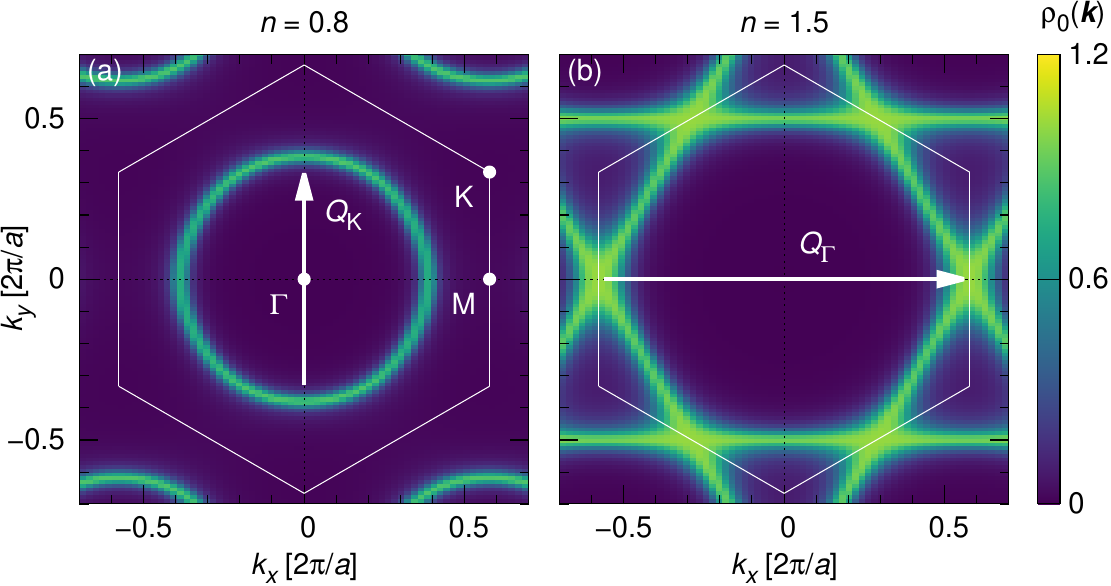}
    \caption{
       Momentum-resolved density of states $\rho_0(\bm{k})$ at the Fermi level in a non-interacting triangular lattice model at $n = 0.8$ and $n = 1.5$ An artificial broadening of width $0.1t$ was applied.
    }
    \label{fig:tri_G0_n0p8-1p5}
\end{figure}

\begin{figure}[hbt]
    \includegraphics[width=\columnwidth]{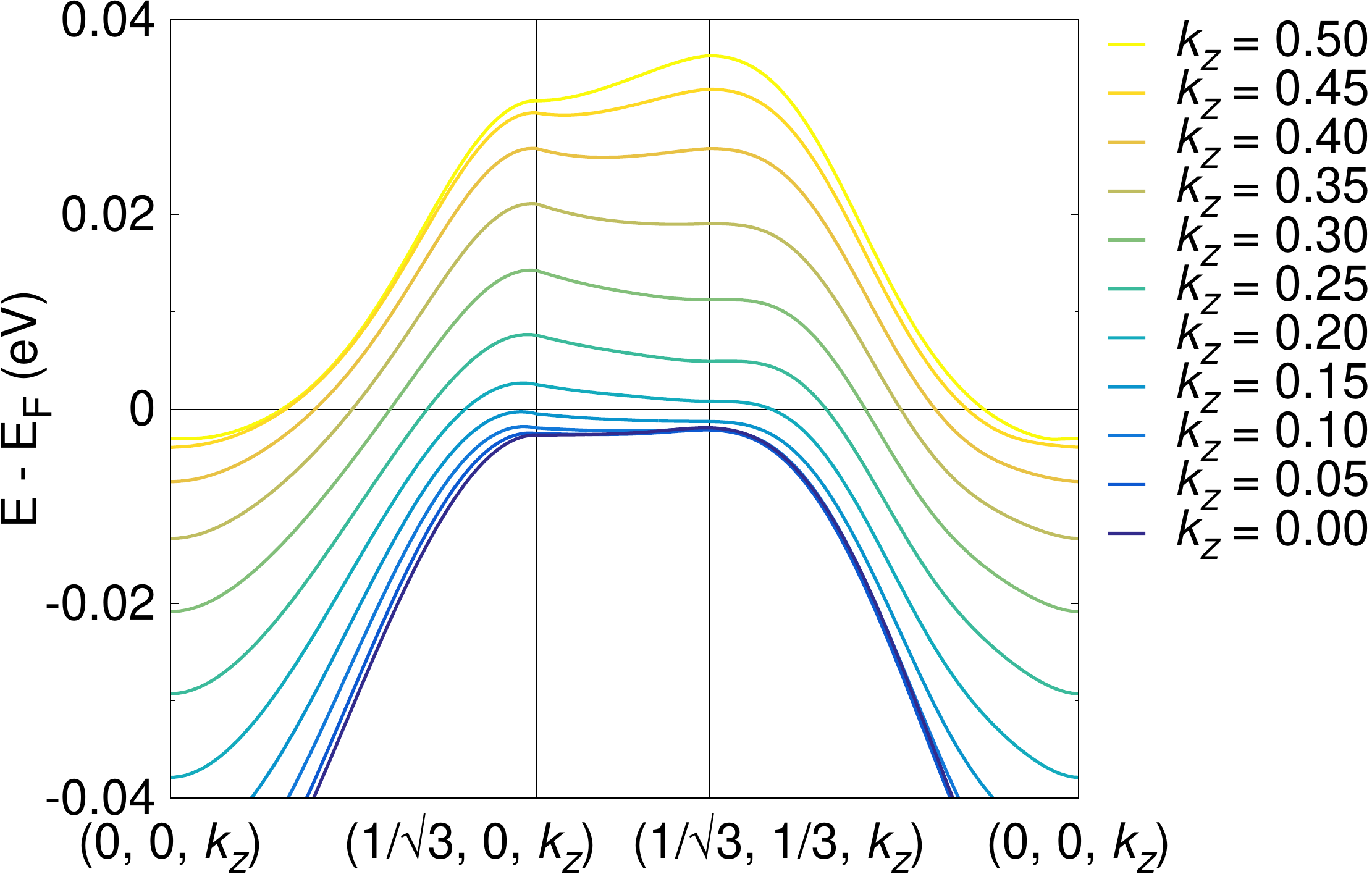}
    \caption{
      Band structure of {\pbcu} along a triangular $k$ path for different values of $k_z$.
      The $x$ and $y$ components are in units of $2\pi/a$,
      and the $z$ component is in units of $2\pi/c$.
      The path at $k_z = 0$ is equivalent to $\Gamma$-M-K-$\Gamma$.
    }
    \label{fig:band_kzvariable}
\end{figure}

\section{Electronic structure}

We perform all electron density functional theory (DFT) calculations for {\pbcu} using the full potential local orbital (FPLO) basis~\cite{Koepernik1999} and the generalized gradient approximation exchange correlation functional~\cite{Perdew1996}. We derive a precise four-band tight binding model using the symmetry conserving projected Wannier functions implemented within FPLO~\cite{Koepernik2023}.

We base our model of the structure of {\pbcu} on the crystal structure of lead apatite {\pba} as determined by Krivovichev and Burns~\cite{Krivovichev1992}. Its $P6_3/m$ (no. 176) space group has an O(4) position that is only occupied by one quarter. We first simplify by choosing one of four symmetry equivalent O(4) positions which reduces the symmetry to $P3$ (no. 143). According to Ref.~\onlinecite{Si2023}, the Pb(2) position with four symmetry equivalent sites is energetically more favorable for Cu substitution. We thus replace one out of four Pb by Cu and perform an internal relaxation of this structure. As DFT structure prediction has well known issues for correlated oxides, we use experimental~\cite{Lee2023a} rather than DFT predicted lattice parameters. The resulting crystal structure is shown in Fig.~\ref{fig:bands}\,(a). It is similar to the structures considered in other works~\cite{Si2023,Jiang2023}.

We show the DFT band structure and density of states of {\pbcu} in Fig.~\ref{fig:bands}\,(b) and (c). Two almost degenerate bands of Cu $3d_{xz}$ and $3d_{yz}$ character are crossing the Fermi level. They are slightly hybridizing with two occupied oxygen bands arising from the extra O in the formula, the O(4) in the {\pba} structure; this has also been observed in Ref.~\cite{Jiang2023}. Therefore, in this study we work with a four-band tight binding model. The upper of the two bands at the Fermi level leads to the three-dimensional Fermi surface shown in Fig.~\ref{fig:3dfs}\,(c). It has mount Fuji shape because the finite $k_z$ dispersion leads to different filling levels of the underlying triangular lattice at every value of $k_z$. Slight hole doping (Fig.~\ref{fig:3dfs}\,(b) and (a)) lead to the rugby ball shape that has been found previously~\cite{Lai2023} and tiny Fermi surfaces from the second Cu $3d$ band. The shape remains qualitatively the same between undoped and moderate electron doping (Fig.~\ref{fig:3dfs}\,(d)). This $k_z$ dispersion of the electronic structure will constitute the main difference to a fully two-dimensional triangular lattice Hamiltonian.

We determine~\cite{Koepernik2023} a four-band tight-binding model
including $3d_{xz}$ and $3d_{yz}$ orbitals of Cu and $2p_x$ and $2p_y$ orbitals of O.
The Hamiltonian $\mathcal{H}_0$ is given by
\begin{equation}
      \mathcal{H}_0 = \sum_{ij\mu\sigma\sigma} t_{ij}^{\mu\nu} c^{\dag}_{i\mu\sigma} c_{j\nu\sigma},
\end{equation}
where $i$ and $j$ are site indices,
$\mu$ and $\nu$ are orbital indices,
and $\sigma$ is the spin index.
The occupation number $n$ is $n = 7$ for the stoichiometric {\pbcu}.
Three bands are fully occupied, and the top band is half-filled without doping and interactions. This means that at stoichiometry, one band forms the Fermi surface.
However, the two upper bands deriving almost entirely from Cu $3d$ are nearly degenerate, and the orbital character of the bands and Fermi surface changes between $3d_{xz}$ and $3d_{yz}$ depending on $k$ direction.

\section{Spin fluctuations}

In order to discuss spin fluctuations, we consider interactions between $3d$ electrons on Cu.
We adopt the usual local interactions, which are represented by the Hamiltonian
\begin{align}
    \mathcal{H}_\mathrm{int}
    &= \sum_{i} \Big[ \frac{U}{2} \sum_{l\sigma} n_{il\sigma} n_{il\bar{\sigma}}
    + \frac{V}{2} \sum_{l \neq m} n_{il} n_{im}
    \notag \\
    &- \frac{J}{2} \sum_{l \neq m} \bm{S}_{il} \cdot \bm{S}_{im}
    + \frac{J'}{2} \sum_{l \neq m} \sum_{\sigma} c^\dag_{il\sigma} c^\dag_{il\sigma} c_{im\bar{\sigma}} c_{im\sigma} \Big],
\end{align}
where
the orbital indices $l$ and $m$ run over two orbitals ($3d_{xz}$ and $3d_{yz}$) out of four.
$U$, $V$, $J$, and $J'$ are intra-orbital Coulomb interaction, inter-orbital Coulomb interaction, Hund's coupling, and pair hopping, respectively.

We apply the FLEX approximation to the four-band Hubbard model. We change the electron number in the range of $6.5 \leq n \leq 7.5$.
We choose spin rotation-invariant interaction parameters~\cite{Guterding2015} $V = U/2$ and $J = J' = U/4$ for the Cu $3d$ orbitals,
and we assume $U = 200\mev$.
We use a $16 \times 16 \times 4$ $\bm{k}$-mesh.


We discuss the momentum-dependent static spin susceptibility, $\chi^\mathrm{s}(\bm{q}) = \sum_{lm} \chi^\mathrm{s}_{lm}(\bm{q})$, in which contributions from $3d_{xz}$ and $3d_{yz}$ orbitals are summed up.
Figure~\ref{fig:chis_n6p7-7p4} shows cuts of $\chi^\mathrm{s}(\bm{q})$ at $q_z=0$, $\pi/2c$, and $\pi/c$ (zone boundary) for $6.7 \leq n \leq 7.4$.
At the formal electron filling of {\pbcu} ($n = 7.0$),
$\chi^\mathrm{s}(\bm{q})$ exhibits peaks at $\bm{q}=\bm{Q}_\mathrm{H}$.
This fluctuation corresponds to an antiferromagnetic (AFM) stacking of the 120$^{\circ}$ state on the triangular lattice.
The AFM fluctuation persists for hole doping.
On the electron doping side, on the other hand, 
the magnetic properties change drastically.
At $n = 7.1$, a peak is present at $\bm{q}=\bm{Q}_\mathrm{\Gamma}$,
which corresponds to the ferromagnetic (FM) fluctuation.
The FM fluctuation becomes dominant for $n \gtrsim 7.2$.

Let us consider the origin of the AFM and FM fluctuations in {\pbcu} in terms of a single-band Hubbard model on a triangular lattice. This model approximately describes the energy dispersion of the top band through $\Gamma$-M-K-$\Gamma$ in Fig.~\ref{fig:bands}. Figure~\ref{fig:tri_chis_n0p4-1p8} shows the spin susceptibility $\chi^\mathrm{s} (\bm{q})$ for several values of the occupation number $n$.
Here, $n$ in this model corresponds to $n-6$ in the four-band model of {\pbcu}.
$\chi^\mathrm{s} (\bm{q})$ exhibits peaks at $\bm{Q}_\mathrm{K}$ around half-filling ($0.6 \leq n \leq 1.0$)
while it has a peak at $\bm{Q}_\mathrm{\Gamma}$ around 3/4-filling ($1.4 \leq n \leq 1.6$). 
These magnetic fluctuations can be understood well with the electronic structure.
Around half-filling, the Fermi surface in the corresponding non-interacting system is a circle around the $\Gamma$ point as shown in Fig.~\ref{fig:tri_G0_n0p8-1p5}\,(a).
The nesting vector agrees with $\bm{Q}_\mathrm{K}$ as indicated by the arrow in Fig.~\ref{fig:tri_G0_n0p8-1p5}\,(a).
On the other hand, the Fermi surface around three-quarter filling is a hexagon inscribed in the first Brillouin zone as shown in Fig.~\ref{fig:tri_G0_n0p8-1p5}\,(b).
The density of states is large at the M point because of the van Hove singularity.
Therefore, dominant particle-hole excitations occur in the $\bm{q}$-vector that connects two M points, that is, $(2\pi/a, 0, 0)$, which corresponds to $\bm{Q}_\mathrm{\Gamma}$.
The Fermi surface thus explains AFM fluctuations corresponding to the 120$^{\circ}$ state for $n\lesssim 1$ and the strong FM fluctuations for $n \simeq 1.5$.

The variations of the magnetic properties between AFM and FM can be understood with the triangular lattice.
However, {\pbcu} shows two unique properties that the triangular lattice does not show.
First, the change from antiferromagnetic to ferromagnetic is very sensitive to electron doping.
It is antiferromagnetic at $n = 7.0$,
and even with very small electron doping, it starts to be ferromagnetic at $n = 7.1$.
0.1 electrons are needed to cause ferromagnetic fluctuations when we start from stoichiometric integer filling.
In the triangular lattice, on the other hand, 0.5 electrons are needed.
Doping 0.5 electrons into a material is usually very difficult in reality.
The second unique property of {\pbcu} is the fact that ferromagnetism persists in the whole region of electron doping while it disappears in the triangular lattice.

These two properties of {\pbcu} result from the three-dimensional Fermi surface.
Even though the highest energy band is almost half-filled in total,
various situations of filling are realized if we look into the two-dimensional $k$-path for different $k_z$: almost empty at $k_z = \pi/c$ and almost fully-filled at $k_z = 0$ as shown in Fig.~\ref{fig:band_kzvariable}.
Therefore, the van Hove singularity at ($1/\sqrt{3}$, 0, $k_z$) appears on the Fermi level even at the stoichiometric filling, and the ferromagnetic fluctuation is enhanced by the mechanism discussed in Fig.~\ref{fig:tri_G0_n0p8-1p5}\,(b).
Moreover, when electrons are doped,
the Fermi level passes the singularity point at a different $k_z$.
This explains the result that the ferromagnetic fluctuation is dominant in a wide range of electron doping levels in {\pbcu}.

\section{Conclusions}

We have studied the consequences of the electronic structure of {\pbcu} for magnetic fluctuations. We find that electronically, {\pbcu} provides an interesting three-quarter filled, two orbital triangular lattice Hamiltonian. As such, it can be partly understood from the one-band triangular lattice Hamiltonian that has previously been studied in the context of organic charge transfer salts and of sodium cobaltate. However, we have worked out two important differences that arise from the three-dimensionality of {\pbcu}, {\it i.e.} from the $k_z$ dispersion of the electronic structure. One consequence is that the change of doping that triggers the transition between antiferromagnetic and ferromagnetic spin fluctuations is much smaller in {\pbcu} compared to the simple triangular lattice. The other consequence is that the ferromagnetic fluctuations in the electron doped region is significantly stabilized in the three-dimensional two-orbital model.

While our FLEX result for the undoped material indicates antiferromagnetic fluctuations, we also find that ferromagnetic fluctuations are reached by very little electron doping. Thus, slightly inhomogeneous samples {\pbx} with $x\approx 1$ would have ferromagnetic as well as antiferromagnetic regions, and this could explain the levitation observed in some of the samples~\cite{Lee2023b,Guo2023,Wu2023} of {\pbcu}.

By focusing on a single site for the position of the doped Cu$^{2+}$ ion, we have introduced a significant simplification, just like in many other studies~\cite{Si2023a,Witt2023,Yue2023,Witt2023}. It is possible that disorder effects will play an important role in {\pbcu}~\cite{Sun2023}. Nevertheless, our study shows that the electronic structure that is present in slightly simplified {\pbcu} has very interesting properties. It is characterized by two degenerate orbitals on a triangular lattice, three-quarter filling and a finite $k_z$ dispersion. As this is a rather generic situation, it could already be realized in other existing materials, or it would not be too hard to design materials with this property. As the hope for superconductivity in {\pbcu} is probably in vain, it makes sense to understand why and possibly find or design a similar material that does support a superconducting ground state. A likely disadvantage of {\pbcu} is the large ratio between interaction strength on the $d$ orbitals and the band width~\cite{Witt2023}; in the search for a better {\pbcu}, wider bands or weaker local interactions would be desirable. If superconductivity could be realized in a dopable material with the electronic structure of {\pbcu}, then we would expect that the order parameter could be tuned between singlet and triplet.

\acknowledgements
We acknowledge useful discussions with S. Kasahara and K. Kobayashi.
The computation in this work has been done using the facilities of the Supercomputer Center, the Institute for Solid State Physics, the University of Tokyo.
M.S.~was supported by JSPS KAKENHI grant No.~22H01181.
J.O.~ was supported by JSPS KAKENHI grant No.~21H01041, No.~21H01003, No.~23H04869.

\bibliography{lk99}

\begin{thebibliography}{32}%
\makeatletter
\providecommand \@ifxundefined [1]{%
 \@ifx{#1\undefined}
}%
\providecommand \@ifnum [1]{%
 \ifnum #1\expandafter \@firstoftwo
 \else \expandafter \@secondoftwo
 \fi
}%
\providecommand \@ifx [1]{%
 \ifx #1\expandafter \@firstoftwo
 \else \expandafter \@secondoftwo
 \fi
}%
\providecommand \natexlab [1]{#1}%
\providecommand \enquote  [1]{``#1''}%
\providecommand \bibnamefont  [1]{#1}%
\providecommand \bibfnamefont [1]{#1}%
\providecommand \citenamefont [1]{#1}%
\providecommand \href@noop [0]{\@secondoftwo}%
\providecommand \href [0]{\begingroup \@sanitize@url \@href}%
\providecommand \@href[1]{\@@startlink{#1}\@@href}%
\providecommand \@@href[1]{\endgroup#1\@@endlink}%
\providecommand \@sanitize@url [0]{\catcode `\\12\catcode `\$12\catcode
  `\&12\catcode `\#12\catcode `\^12\catcode `\_12\catcode `\%12\relax}%
\providecommand \@@startlink[1]{}%
\providecommand \@@endlink[0]{}%
\providecommand \url  [0]{\begingroup\@sanitize@url \@url }%
\providecommand \@url [1]{\endgroup\@href {#1}{\urlprefix }}%
\providecommand \urlprefix  [0]{URL }%
\providecommand \Eprint [0]{\href }%
\providecommand \doibase [0]{https://doi.org/}%
\providecommand \selectlanguage [0]{\@gobble}%
\providecommand \bibinfo  [0]{\@secondoftwo}%
\providecommand \bibfield  [0]{\@secondoftwo}%
\providecommand \translation [1]{[#1]}%
\providecommand \BibitemOpen [0]{}%
\providecommand \bibitemStop [0]{}%
\providecommand \bibitemNoStop [0]{.\EOS\space}%
\providecommand \EOS [0]{\spacefactor3000\relax}%
\providecommand \BibitemShut  [1]{\csname bibitem#1\endcsname}%
\let\auto@bib@innerbib\@empty
\bibitem [{\citenamefont {Lee}\ \emph {et~al.}(2023{\natexlab{a}})\citenamefont
  {Lee}, \citenamefont {Kim},\ and\ \citenamefont {Kwon}}]{Lee2023a}%
  \BibitemOpen
  \bibfield  {author} {\bibinfo {author} {\bibfnamefont {S.}~\bibnamefont
  {Lee}}, \bibinfo {author} {\bibfnamefont {J.-H.}\ \bibnamefont {Kim}},\ and\
  \bibinfo {author} {\bibfnamefont {Y.-W.}\ \bibnamefont {Kwon}},\ }\href@noop
  {} {\bibinfo {title} {The first room-temperature ambient-pressure
  superconductor}} (\bibinfo {year} {2023}{\natexlab{a}}),\ \Eprint
  {https://arxiv.org/abs/2307.12008} {arXiv:2307.12008 [cond-mat.supr-con]}
  \BibitemShut {NoStop}%
\bibitem [{\citenamefont {Lee}\ \emph {et~al.}(2023{\natexlab{b}})\citenamefont
  {Lee}, \citenamefont {Kim}, \citenamefont {Kim}, \citenamefont {Im},
  \citenamefont {An},\ and\ \citenamefont {Auh}}]{Lee2023b}%
  \BibitemOpen
  \bibfield  {author} {\bibinfo {author} {\bibfnamefont {S.}~\bibnamefont
  {Lee}}, \bibinfo {author} {\bibfnamefont {J.}~\bibnamefont {Kim}}, \bibinfo
  {author} {\bibfnamefont {H.-T.}\ \bibnamefont {Kim}}, \bibinfo {author}
  {\bibfnamefont {S.}~\bibnamefont {Im}}, \bibinfo {author} {\bibfnamefont
  {S.}~\bibnamefont {An}},\ and\ \bibinfo {author} {\bibfnamefont {K.~H.}\
  \bibnamefont {Auh}},\ }\href@noop {} {\bibinfo {title} {Superconductor
  \ce{Pb$_{10-x}$Cu$_x$(PO4)6O} showing levitation at room temperature and
  atmospheric pressure and mechanism}} (\bibinfo {year} {2023}{\natexlab{b}}),\
  \Eprint {https://arxiv.org/abs/2307.12037} {arXiv:2307.12037
  [cond-mat.supr-con]} \BibitemShut {NoStop}%
\bibitem [{\citenamefont {Kumar}\ \emph
  {et~al.}(2023{\natexlab{a}})\citenamefont {Kumar}, \citenamefont {Karn},\
  and\ \citenamefont {Awana}}]{Kumar2023}%
  \BibitemOpen
  \bibfield  {author} {\bibinfo {author} {\bibfnamefont {K.}~\bibnamefont
  {Kumar}}, \bibinfo {author} {\bibfnamefont {N.~K.}\ \bibnamefont {Karn}},\
  and\ \bibinfo {author} {\bibfnamefont {V.~P.~S.}\ \bibnamefont {Awana}},\
  }\href@noop {} {\bibinfo {title} {Synthesis of possible room temperature
  superconductor {LK}-99: \ce{Pb9Cu(PO4)6O}}} (\bibinfo {year}
  {2023}{\natexlab{a}}),\ \Eprint {https://arxiv.org/abs/2307.16402}
  {arXiv:2307.16402 [cond-mat.supr-con]} \BibitemShut {NoStop}%
\bibitem [{\citenamefont {Hou}\ \emph {et~al.}(2023)\citenamefont {Hou},
  \citenamefont {Wei}, \citenamefont {Zhou}, \citenamefont {Sun},\ and\
  \citenamefont {Shi}}]{Hou2023}%
  \BibitemOpen
  \bibfield  {author} {\bibinfo {author} {\bibfnamefont {Q.}~\bibnamefont
  {Hou}}, \bibinfo {author} {\bibfnamefont {W.}~\bibnamefont {Wei}}, \bibinfo
  {author} {\bibfnamefont {X.}~\bibnamefont {Zhou}}, \bibinfo {author}
  {\bibfnamefont {Y.}~\bibnamefont {Sun}},\ and\ \bibinfo {author}
  {\bibfnamefont {Z.}~\bibnamefont {Shi}},\ }\href@noop {} {\bibinfo {title}
  {Observation of zero resistance above 100$^\circ$ {K} in
  \ce{Pb$_{10-x}$Cu$_x$(PO4)6O}}} (\bibinfo {year} {2023}),\ \Eprint
  {https://arxiv.org/abs/2308.01192} {arXiv:2308.01192 [cond-mat.supr-con]}
  \BibitemShut {NoStop}%
\bibitem [{\citenamefont {Guo}\ \emph {et~al.}(2023)\citenamefont {Guo},
  \citenamefont {Li},\ and\ \citenamefont {Jia}}]{Guo2023}%
  \BibitemOpen
  \bibfield  {author} {\bibinfo {author} {\bibfnamefont {K.}~\bibnamefont
  {Guo}}, \bibinfo {author} {\bibfnamefont {Y.}~\bibnamefont {Li}},\ and\
  \bibinfo {author} {\bibfnamefont {S.}~\bibnamefont {Jia}},\ }\href@noop {}
  {\bibinfo {title} {Ferromagnetic half levitation of {LK}-99-like synthetic
  samples}} (\bibinfo {year} {2023}),\ \Eprint
  {https://arxiv.org/abs/2308.03110} {arXiv:2308.03110 [cond-mat.supr-con]}
  \BibitemShut {NoStop}%
\bibitem [{\citenamefont {Kumar}\ \emph
  {et~al.}(2023{\natexlab{b}})\citenamefont {Kumar}, \citenamefont {Karn},
  \citenamefont {Kumar},\ and\ \citenamefont {Awana}}]{Kumar2023a}%
  \BibitemOpen
  \bibfield  {author} {\bibinfo {author} {\bibfnamefont {K.}~\bibnamefont
  {Kumar}}, \bibinfo {author} {\bibfnamefont {N.~K.}\ \bibnamefont {Karn}},
  \bibinfo {author} {\bibfnamefont {Y.}~\bibnamefont {Kumar}},\ and\ \bibinfo
  {author} {\bibfnamefont {V.~P.~S.}\ \bibnamefont {Awana}},\ }\href@noop {}
  {\bibinfo {title} {Absence of superconductivity in {LK}-99 at ambient
  conditions}} (\bibinfo {year} {2023}{\natexlab{b}}),\ \Eprint
  {https://arxiv.org/abs/2308.03544} {arXiv:2308.03544 [cond-mat.supr-con]}
  \BibitemShut {NoStop}%
\bibitem [{\citenamefont {Jain}(2023)}]{Jain2023}%
  \BibitemOpen
  \bibfield  {author} {\bibinfo {author} {\bibfnamefont {P.~K.}\ \bibnamefont
  {Jain}},\ }\href@noop {} {\bibinfo {title} {Phase transition of copper (i)
  sulfide and its implication for purported superconductivity of {LK}-99}}
  (\bibinfo {year} {2023}),\ \Eprint {https://arxiv.org/abs/2308.05222}
  {arXiv:2308.05222 [cond-mat.supr-con]} \BibitemShut {NoStop}%
\bibitem [{\citenamefont {Zhu}\ \emph {et~al.}(2023)\citenamefont {Zhu},
  \citenamefont {Wu}, \citenamefont {Li},\ and\ \citenamefont {Luo}}]{Zhu2023}%
  \BibitemOpen
  \bibfield  {author} {\bibinfo {author} {\bibfnamefont {S.}~\bibnamefont
  {Zhu}}, \bibinfo {author} {\bibfnamefont {W.}~\bibnamefont {Wu}}, \bibinfo
  {author} {\bibfnamefont {Z.}~\bibnamefont {Li}},\ and\ \bibinfo {author}
  {\bibfnamefont {J.}~\bibnamefont {Luo}},\ }\href@noop {} {\bibinfo {title}
  {First order transition in \ce{Pb$_{10-x}$Cu$_x$(PO4)6O} ($0.9<x<1.1$)
  containing \ce{Cu2S}}} (\bibinfo {year} {2023}),\ \Eprint
  {https://arxiv.org/abs/2308.04353} {arXiv:2308.04353 [cond-mat.supr-con]}
  \BibitemShut {NoStop}%
\bibitem [{\citenamefont {Wu}\ \emph {et~al.}(2023)\citenamefont {Wu},
  \citenamefont {Yang}, \citenamefont {Xiao},\ and\ \citenamefont
  {Chang}}]{Wu2023}%
  \BibitemOpen
  \bibfield  {author} {\bibinfo {author} {\bibfnamefont {H.}~\bibnamefont
  {Wu}}, \bibinfo {author} {\bibfnamefont {L.}~\bibnamefont {Yang}}, \bibinfo
  {author} {\bibfnamefont {B.}~\bibnamefont {Xiao}},\ and\ \bibinfo {author}
  {\bibfnamefont {H.}~\bibnamefont {Chang}},\ }\href@noop {} {\bibinfo {title}
  {Successful growth and room temperature ambient-pressure magnetic levitation
  of {LK}-99}} (\bibinfo {year} {2023}),\ \Eprint
  {https://arxiv.org/abs/2308.01516} {arXiv:2308.01516 [cond-mat.supr-con]}
  \BibitemShut {NoStop}%
\bibitem [{\citenamefont {Puphal}\ \emph {et~al.}(2023)\citenamefont {Puphal},
  \citenamefont {Akbar}, \citenamefont {Hepting}, \citenamefont {Goering},
  \citenamefont {Isobe}, \citenamefont {Nugroho},\ and\ \citenamefont
  {Keimer}}]{Puphal2023}%
  \BibitemOpen
  \bibfield  {author} {\bibinfo {author} {\bibfnamefont {P.}~\bibnamefont
  {Puphal}}, \bibinfo {author} {\bibfnamefont {M.~Y.~P.}\ \bibnamefont
  {Akbar}}, \bibinfo {author} {\bibfnamefont {M.}~\bibnamefont {Hepting}},
  \bibinfo {author} {\bibfnamefont {E.}~\bibnamefont {Goering}}, \bibinfo
  {author} {\bibfnamefont {M.}~\bibnamefont {Isobe}}, \bibinfo {author}
  {\bibfnamefont {A.~A.}\ \bibnamefont {Nugroho}},\ and\ \bibinfo {author}
  {\bibfnamefont {B.}~\bibnamefont {Keimer}},\ }\href@noop {} {\bibinfo {title}
  {Single crystal synthesis, structure, and magnetism of
  \ce{Pb$_{10-x}$Cu$_x$(PO4)6O}}} (\bibinfo {year} {2023}),\ \Eprint
  {https://arxiv.org/abs/2308.06256} {arXiv:2308.06256 [cond-mat.supr-con]}
  \BibitemShut {NoStop}%
\bibitem [{\citenamefont {Lai}\ \emph {et~al.}(2024)\citenamefont {Lai},
  \citenamefont {Li}, \citenamefont {Liu}, \citenamefont {Sun},\ and\
  \citenamefont {Chen}}]{Lai2023}%
  \BibitemOpen
  \bibfield  {author} {\bibinfo {author} {\bibfnamefont {J.}~\bibnamefont
  {Lai}}, \bibinfo {author} {\bibfnamefont {J.}~\bibnamefont {Li}}, \bibinfo
  {author} {\bibfnamefont {P.}~\bibnamefont {Liu}}, \bibinfo {author}
  {\bibfnamefont {Y.}~\bibnamefont {Sun}},\ and\ \bibinfo {author}
  {\bibfnamefont {X.-Q.}\ \bibnamefont {Chen}},\ }\bibfield  {title} {\bibinfo
  {title} {First-principles study on the electronic structure of
  \ce{Pb$_{10-x}$Cu$_x$(PO4)6O} ($x=0$, 1)},\ }\href@noop {} {\bibfield
  {journal} {\bibinfo  {journal} {J. Mater. Sci. Technol.}\ }\textbf {\bibinfo
  {volume} {171}},\ \bibinfo {pages} {66} (\bibinfo {year} {2024})}\BibitemShut
  {NoStop}%
\bibitem [{\citenamefont {Griffin}(2023)}]{Griffin2023}%
  \BibitemOpen
  \bibfield  {author} {\bibinfo {author} {\bibfnamefont {S.~M.}\ \bibnamefont
  {Griffin}},\ }\href@noop {} {\bibinfo {title} {Origin of correlated isolated
  flat bands in copper-substituted lead phosphate apatite}} (\bibinfo {year}
  {2023}),\ \Eprint {https://arxiv.org/abs/2307.16892} {arXiv:2307.16892
  [cond-mat.supr-con]} \BibitemShut {NoStop}%
\bibitem [{\citenamefont {Si}\ and\ \citenamefont {Held}(2023)}]{Si2023}%
  \BibitemOpen
  \bibfield  {author} {\bibinfo {author} {\bibfnamefont {L.}~\bibnamefont
  {Si}}\ and\ \bibinfo {author} {\bibfnamefont {K.}~\bibnamefont {Held}},\
  }\href@noop {} {\bibinfo {title} {Electronic structure of the putative
  room-temperature superconductor \ce{Pb9Cu(PO4)6O}}} (\bibinfo {year}
  {2023}),\ \Eprint {https://arxiv.org/abs/2308.00676} {arXiv:2308.00676
  [cond-mat.supr-con]} \BibitemShut {NoStop}%
\bibitem [{\citenamefont {Kurleto}\ \emph {et~al.}(2023)\citenamefont
  {Kurleto}, \citenamefont {Lany}, \citenamefont {Pashov}, \citenamefont
  {Acharya}, \citenamefont {van Schilfgaarde},\ and\ \citenamefont
  {Dessau}}]{Kurleto2023}%
  \BibitemOpen
  \bibfield  {author} {\bibinfo {author} {\bibfnamefont {R.}~\bibnamefont
  {Kurleto}}, \bibinfo {author} {\bibfnamefont {S.}~\bibnamefont {Lany}},
  \bibinfo {author} {\bibfnamefont {D.}~\bibnamefont {Pashov}}, \bibinfo
  {author} {\bibfnamefont {S.}~\bibnamefont {Acharya}}, \bibinfo {author}
  {\bibfnamefont {M.}~\bibnamefont {van Schilfgaarde}},\ and\ \bibinfo {author}
  {\bibfnamefont {D.~S.}\ \bibnamefont {Dessau}},\ }\href@noop {} {\bibinfo
  {title} {Pb-apatite framework as a generator of novel flat-band \ce{CuO}
  based physics, including possible room temperature superconductivity}}
  (\bibinfo {year} {2023}),\ \Eprint {https://arxiv.org/abs/2308.00698}
  {arXiv:2308.00698 [cond-mat.supr-con]} \BibitemShut {NoStop}%
\bibitem [{\citenamefont {Cabezas-Escares}\ \emph {et~al.}(2023)\citenamefont
  {Cabezas-Escares}, \citenamefont {Barrera}, \citenamefont {Cardenas},\ and\
  \citenamefont {Munoz}}]{Cabezas2023}%
  \BibitemOpen
  \bibfield  {author} {\bibinfo {author} {\bibfnamefont {J.}~\bibnamefont
  {Cabezas-Escares}}, \bibinfo {author} {\bibfnamefont {N.~F.}\ \bibnamefont
  {Barrera}}, \bibinfo {author} {\bibfnamefont {C.}~\bibnamefont {Cardenas}},\
  and\ \bibinfo {author} {\bibfnamefont {F.}~\bibnamefont {Munoz}},\
  }\href@noop {} {\bibinfo {title} {Theoretical insight on the {LK}-99
  material}} (\bibinfo {year} {2023}),\ \Eprint
  {https://arxiv.org/abs/2308.01135} {arXiv:2308.01135 [cond-mat.supr-con]}
  \BibitemShut {NoStop}%
\bibitem [{\citenamefont {Tavakol}\ and\ \citenamefont
  {Scaffidi}(2023)}]{Tavakol2023}%
  \BibitemOpen
  \bibfield  {author} {\bibinfo {author} {\bibfnamefont {O.}~\bibnamefont
  {Tavakol}}\ and\ \bibinfo {author} {\bibfnamefont {T.}~\bibnamefont
  {Scaffidi}},\ }\href@noop {} {\bibinfo {title} {Minimal model for the flat
  bands in copper-substituted lead phosphate apatite}} (\bibinfo {year}
  {2023}),\ \Eprint {https://arxiv.org/abs/2308.01315} {arXiv:2308.01315
  [cond-mat.supr-con]} \BibitemShut {NoStop}%
\bibitem [{\citenamefont {Lee}\ and\ \citenamefont {Dai}(2023)}]{Lee2023c}%
  \BibitemOpen
  \bibfield  {author} {\bibinfo {author} {\bibfnamefont {P.~A.}\ \bibnamefont
  {Lee}}\ and\ \bibinfo {author} {\bibfnamefont {Z.}~\bibnamefont {Dai}},\
  }\href@noop {} {\bibinfo {title} {Effective model for pb$_9$cu(po$_4$)$_6$o}}
  (\bibinfo {year} {2023}),\ \Eprint {https://arxiv.org/abs/2308.04480}
  {arXiv:2308.04480 [cond-mat.supr-con]} \BibitemShut {NoStop}%
\bibitem [{\citenamefont {Hirschmann}\ and\ \citenamefont
  {Mitscherling}(2023)}]{Hirschmann2023}%
  \BibitemOpen
  \bibfield  {author} {\bibinfo {author} {\bibfnamefont {M.~M.}\ \bibnamefont
  {Hirschmann}}\ and\ \bibinfo {author} {\bibfnamefont {J.}~\bibnamefont
  {Mitscherling}},\ }\href@noop {} {\bibinfo {title} {Tight-binding models for
  {SG} 143 ({P}3) and application to recent {DFT} results on copper-doped lead
  apatite}} (\bibinfo {year} {2023}),\ \Eprint
  {https://arxiv.org/abs/2308.03751} {arXiv:2308.03751 [cond-mat.mes-hall]}
  \BibitemShut {NoStop}%
\bibitem [{\citenamefont {Korotin}\ \emph {et~al.}(2023)\citenamefont
  {Korotin}, \citenamefont {Novoselov}, \citenamefont {Shorikov}, \citenamefont
  {Anisimov},\ and\ \citenamefont {Oganov}}]{Korotin2023}%
  \BibitemOpen
  \bibfield  {author} {\bibinfo {author} {\bibfnamefont {D.~M.}\ \bibnamefont
  {Korotin}}, \bibinfo {author} {\bibfnamefont {D.~Y.}\ \bibnamefont
  {Novoselov}}, \bibinfo {author} {\bibfnamefont {A.~O.}\ \bibnamefont
  {Shorikov}}, \bibinfo {author} {\bibfnamefont {V.~I.}\ \bibnamefont
  {Anisimov}},\ and\ \bibinfo {author} {\bibfnamefont {A.~R.}\ \bibnamefont
  {Oganov}},\ }\href@noop {} {\bibinfo {title} {Electronic correlations in
  promising room-temperature superconductor \ce{Pb9Cu(PO4)6O}: a {DFT}+{DMFT}
  study}} (\bibinfo {year} {2023}),\ \Eprint {https://arxiv.org/abs/2308.04301}
  {arXiv:2308.04301 [cond-mat.supr-con]} \BibitemShut {NoStop}%
\bibitem [{\citenamefont {Si}\ \emph {et~al.}(2023)\citenamefont {Si},
  \citenamefont {Wallerberger}, \citenamefont {Smolyanyuk}, \citenamefont
  {di~Cataldo}, \citenamefont {Tomczak},\ and\ \citenamefont {Held}}]{Si2023a}%
  \BibitemOpen
  \bibfield  {author} {\bibinfo {author} {\bibfnamefont {L.}~\bibnamefont
  {Si}}, \bibinfo {author} {\bibfnamefont {M.}~\bibnamefont {Wallerberger}},
  \bibinfo {author} {\bibfnamefont {A.}~\bibnamefont {Smolyanyuk}}, \bibinfo
  {author} {\bibfnamefont {S.}~\bibnamefont {di~Cataldo}}, \bibinfo {author}
  {\bibfnamefont {J.~M.}\ \bibnamefont {Tomczak}},\ and\ \bibinfo {author}
  {\bibfnamefont {K.}~\bibnamefont {Held}},\ }\href@noop {} {\bibinfo {title}
  {\ce{Pb$_{10-x}$Cu$_x$(PO4)6O}: a mott or charge transfer insulator in need
  of further doping for (super)conductivity}} (\bibinfo {year} {2023}),\
  \Eprint {https://arxiv.org/abs/2308.04427} {arXiv:2308.04427
  [cond-mat.supr-con]} \BibitemShut {NoStop}%
\bibitem [{\citenamefont {Yue}\ \emph {et~al.}(2023)\citenamefont {Yue},
  \citenamefont {Christiansson},\ and\ \citenamefont {Werner}}]{Yue2023}%
  \BibitemOpen
  \bibfield  {author} {\bibinfo {author} {\bibfnamefont {C.}~\bibnamefont
  {Yue}}, \bibinfo {author} {\bibfnamefont {V.}~\bibnamefont {Christiansson}},\
  and\ \bibinfo {author} {\bibfnamefont {P.}~\bibnamefont {Werner}},\
  }\href@noop {} {\bibinfo {title} {Correlated electronic structure of
  \ce{Pb$_{10-x}$Cu$_x$(PO4)6O}}} (\bibinfo {year} {2023}),\ \Eprint
  {https://arxiv.org/abs/2308.04976} {arXiv:2308.04976 [cond-mat.str-el]}
  \BibitemShut {NoStop}%
\bibitem [{\citenamefont {Oh}\ and\ \citenamefont {Zhang}(2023)}]{Oh2023}%
  \BibitemOpen
  \bibfield  {author} {\bibinfo {author} {\bibfnamefont {H.}~\bibnamefont
  {Oh}}\ and\ \bibinfo {author} {\bibfnamefont {Y.-H.}\ \bibnamefont {Zhang}},\
  }\href@noop {} {\bibinfo {title} {S-wave pairing in a two-orbital t-{J} model
  on triangular lattice: possible application to
  \ce{Pb$_{10-x}$Cu$_x$(PO4)6O}}} (\bibinfo {year} {2023}),\ \Eprint
  {https://arxiv.org/abs/2308.02469} {arXiv:2308.02469 [cond-mat.supr-con]}
  \BibitemShut {NoStop}%
\bibitem [{\citenamefont {Witt}\ \emph {et~al.}(2023)\citenamefont {Witt},
  \citenamefont {Si}, \citenamefont {Tomczak}, \citenamefont {Held},\ and\
  \citenamefont {Wehling}}]{Witt2023}%
  \BibitemOpen
  \bibfield  {author} {\bibinfo {author} {\bibfnamefont {N.}~\bibnamefont
  {Witt}}, \bibinfo {author} {\bibfnamefont {L.}~\bibnamefont {Si}}, \bibinfo
  {author} {\bibfnamefont {J.~M.}\ \bibnamefont {Tomczak}}, \bibinfo {author}
  {\bibfnamefont {K.}~\bibnamefont {Held}},\ and\ \bibinfo {author}
  {\bibfnamefont {T.}~\bibnamefont {Wehling}},\ }\href@noop {} {\bibinfo
  {title} {No superconductivity in \ce{Pb9Cu1(PO4)6O} found in orbital and spin
  fluctuation exchange calculations}} (\bibinfo {year} {2023}),\ \Eprint
  {https://arxiv.org/abs/2308.07261} {arXiv:2308.07261 [cond-mat.supr-con]}
  \BibitemShut {NoStop}%
\bibitem [{\citenamefont {Sun}\ \emph {et~al.}(2023)\citenamefont {Sun},
  \citenamefont {Ho},\ and\ \citenamefont {Antropov}}]{Sun2023}%
  \BibitemOpen
  \bibfield  {author} {\bibinfo {author} {\bibfnamefont {Y.}~\bibnamefont
  {Sun}}, \bibinfo {author} {\bibfnamefont {K.-M.}\ \bibnamefont {Ho}},\ and\
  \bibinfo {author} {\bibfnamefont {V.}~\bibnamefont {Antropov}},\ }\href@noop
  {} {\bibinfo {title} {Metallization and spin fluctuations in {C}u-doped lead
  apatite}} (\bibinfo {year} {2023}),\ \Eprint
  {https://arxiv.org/abs/2308.03454} {arXiv:2308.03454 [cond-mat.supr-con]}
  \BibitemShut {NoStop}%
\bibitem [{\citenamefont {Bickers}\ \emph {et~al.}(1989)\citenamefont
  {Bickers}, \citenamefont {Scalapino},\ and\ \citenamefont
  {White}}]{Bickers1989}%
  \BibitemOpen
  \bibfield  {author} {\bibinfo {author} {\bibfnamefont {N.~E.}\ \bibnamefont
  {Bickers}}, \bibinfo {author} {\bibfnamefont {D.~J.}\ \bibnamefont
  {Scalapino}},\ and\ \bibinfo {author} {\bibfnamefont {S.~R.}\ \bibnamefont
  {White}},\ }\bibfield  {title} {\bibinfo {title} {{Conserving Approximations
  for Strongly Correlated Electron Systems: Bethe-Salpeter Equation and
  Dynamics for the Two-Dimensional Hubbard Model}},\ }\href
  {https://doi.org/10.1103/PhysRevLett.62.961} {\bibfield  {journal} {\bibinfo
  {journal} {Phys. Rev. Lett.}\ }\textbf {\bibinfo {volume} {62}},\ \bibinfo
  {pages} {961} (\bibinfo {year} {1989})}\BibitemShut {NoStop}%
\bibitem [{\citenamefont {Ikeda}\ \emph {et~al.}(2010)\citenamefont {Ikeda},
  \citenamefont {Arita},\ and\ \citenamefont {Kune{\v s}}}]{Ikeda2010}%
  \BibitemOpen
  \bibfield  {author} {\bibinfo {author} {\bibfnamefont {H.}~\bibnamefont
  {Ikeda}}, \bibinfo {author} {\bibfnamefont {R.}~\bibnamefont {Arita}},\ and\
  \bibinfo {author} {\bibfnamefont {J.}~\bibnamefont {Kune{\v s}}},\ }\bibfield
   {title} {\bibinfo {title} {Phase diagram and gap anisotropy in iron-pnictide
  superconductors},\ }\href
  {https://journals.aps.org/prb/abstract/10.1103/PhysRevB.81.054502} {\bibfield
   {journal} {\bibinfo  {journal} {Phys. Rev. B Condens. Matter}\ }\textbf
  {\bibinfo {volume} {81}},\ \bibinfo {pages} {054502} (\bibinfo {year}
  {2010})}\BibitemShut {NoStop}%
\bibitem [{\citenamefont {Koepernik}\ and\ \citenamefont
  {Eschrig}(1999)}]{Koepernik1999}%
  \BibitemOpen
  \bibfield  {author} {\bibinfo {author} {\bibfnamefont {K.}~\bibnamefont
  {Koepernik}}\ and\ \bibinfo {author} {\bibfnamefont {H.}~\bibnamefont
  {Eschrig}},\ }\bibfield  {title} {\bibinfo {title} {Full-potential
  nonorthogonal local-orbital minimum-basis band-structure scheme},\ }\href
  {https://doi.org/10.1103/PhysRevB.59.1743} {\bibfield  {journal} {\bibinfo
  {journal} {Phys. Rev. B}\ }\textbf {\bibinfo {volume} {59}},\ \bibinfo
  {pages} {1743} (\bibinfo {year} {1999})}\BibitemShut {NoStop}%
\bibitem [{\citenamefont {Perdew}\ \emph {et~al.}(1996)\citenamefont {Perdew},
  \citenamefont {Burke},\ and\ \citenamefont {Ernzerhof}}]{Perdew1996}%
  \BibitemOpen
  \bibfield  {author} {\bibinfo {author} {\bibfnamefont {J.~P.}\ \bibnamefont
  {Perdew}}, \bibinfo {author} {\bibfnamefont {K.}~\bibnamefont {Burke}},\ and\
  \bibinfo {author} {\bibfnamefont {M.}~\bibnamefont {Ernzerhof}},\ }\bibfield
  {title} {\bibinfo {title} {Generalized gradient approximation made simple},\
  }\href {https://doi.org/10.1103/PhysRevLett.77.3865} {\bibfield  {journal}
  {\bibinfo  {journal} {Phys. Rev. Lett.}\ }\textbf {\bibinfo {volume} {77}},\
  \bibinfo {pages} {3865} (\bibinfo {year} {1996})}\BibitemShut {NoStop}%
\bibitem [{\citenamefont {Koepernik}\ \emph {et~al.}(2023)\citenamefont
  {Koepernik}, \citenamefont {Janson}, \citenamefont {Sun},\ and\ \citenamefont
  {van~den Brink}}]{Koepernik2023}%
  \BibitemOpen
  \bibfield  {author} {\bibinfo {author} {\bibfnamefont {K.}~\bibnamefont
  {Koepernik}}, \bibinfo {author} {\bibfnamefont {O.}~\bibnamefont {Janson}},
  \bibinfo {author} {\bibfnamefont {Y.}~\bibnamefont {Sun}},\ and\ \bibinfo
  {author} {\bibfnamefont {J.}~\bibnamefont {van~den Brink}},\ }\bibfield
  {title} {\bibinfo {title} {Symmetry-conserving maximally projected {W}annier
  functions},\ }\href {https://doi.org/10.1103/PhysRevB.107.235135} {\bibfield
  {journal} {\bibinfo  {journal} {Phys. Rev. B}\ }\textbf {\bibinfo {volume}
  {107}},\ \bibinfo {pages} {235135} (\bibinfo {year} {2023})}\BibitemShut
  {NoStop}%
\bibitem [{\citenamefont {Krivovichev}\ and\ \citenamefont
  {Burns}(2003)}]{Krivovichev1992}%
  \BibitemOpen
  \bibfield  {author} {\bibinfo {author} {\bibfnamefont {S.~V.}\ \bibnamefont
  {Krivovichev}}\ and\ \bibinfo {author} {\bibfnamefont {P.~C.}\ \bibnamefont
  {Burns}},\ }\bibfield  {title} {\bibinfo {title} {Crystal chemistry of lead
  oxide phosphates: crystal structures of \ce{Pb4O(PO4)2}, \ce{Pb8O5(PO4)2} and
  \ce{Pb10(PO4)6O}},\ }\href {https://doi.org/doi:10.1524/zkri.218.5.357.20732}
  {\bibfield  {journal} {\bibinfo  {journal} {Z. Kristallogr.}\ }\textbf
  {\bibinfo {volume} {218}},\ \bibinfo {pages} {357} (\bibinfo {year}
  {2003})}\BibitemShut {NoStop}%
\bibitem [{\citenamefont {Jiang}\ \emph {et~al.}(2023)\citenamefont {Jiang},
  \citenamefont {Lee}, \citenamefont {Herzog-Arbeitman}, \citenamefont {Yu},
  \citenamefont {Feng}, \citenamefont {Hu}, \citenamefont {Călugăru},
  \citenamefont {Brodale}, \citenamefont {Gormley}, \citenamefont {Vergniory},
  \citenamefont {Felser}, \citenamefont {Blanco-Canosa}, \citenamefont
  {Hendon}, \citenamefont {Schoop},\ and\ \citenamefont
  {Bernevig}}]{Jiang2023}%
  \BibitemOpen
  \bibfield  {author} {\bibinfo {author} {\bibfnamefont {Y.}~\bibnamefont
  {Jiang}}, \bibinfo {author} {\bibfnamefont {S.~B.}\ \bibnamefont {Lee}},
  \bibinfo {author} {\bibfnamefont {J.}~\bibnamefont {Herzog-Arbeitman}},
  \bibinfo {author} {\bibfnamefont {J.}~\bibnamefont {Yu}}, \bibinfo {author}
  {\bibfnamefont {X.}~\bibnamefont {Feng}}, \bibinfo {author} {\bibfnamefont
  {H.}~\bibnamefont {Hu}}, \bibinfo {author} {\bibfnamefont {D.}~\bibnamefont
  {Călugăru}}, \bibinfo {author} {\bibfnamefont {P.~S.}\ \bibnamefont
  {Brodale}}, \bibinfo {author} {\bibfnamefont {E.~L.}\ \bibnamefont
  {Gormley}}, \bibinfo {author} {\bibfnamefont {M.~G.}\ \bibnamefont
  {Vergniory}}, \bibinfo {author} {\bibfnamefont {C.}~\bibnamefont {Felser}},
  \bibinfo {author} {\bibfnamefont {S.}~\bibnamefont {Blanco-Canosa}}, \bibinfo
  {author} {\bibfnamefont {C.~H.}\ \bibnamefont {Hendon}}, \bibinfo {author}
  {\bibfnamefont {L.~M.}\ \bibnamefont {Schoop}},\ and\ \bibinfo {author}
  {\bibfnamefont {B.~A.}\ \bibnamefont {Bernevig}},\ }\href@noop {} {\bibinfo
  {title} {\ce{Pb9Cu(PO4)6(OH)2}: Phonon bands, localized flat band magnetism,
  models, and chemical analysis}} (\bibinfo {year} {2023}),\ \Eprint
  {https://arxiv.org/abs/2308.05143} {arXiv:2308.05143 [cond-mat.supr-con]}
  \BibitemShut {NoStop}%
\bibitem [{\citenamefont {Guterding}\ \emph {et~al.}(2015)\citenamefont
  {Guterding}, \citenamefont {Jeschke}, \citenamefont {Hirschfeld},\ and\
  \citenamefont {Valent\'{\i}}}]{Guterding2015}%
  \BibitemOpen
  \bibfield  {author} {\bibinfo {author} {\bibfnamefont {D.}~\bibnamefont
  {Guterding}}, \bibinfo {author} {\bibfnamefont {H.~O.}\ \bibnamefont
  {Jeschke}}, \bibinfo {author} {\bibfnamefont {P.~J.}\ \bibnamefont
  {Hirschfeld}},\ and\ \bibinfo {author} {\bibfnamefont {R.}~\bibnamefont
  {Valent\'{\i}}},\ }\bibfield  {title} {\bibinfo {title} {{Unified picture of
  the doping dependence of superconducting transition temperatures in alkali
  metal/ammonia intercalated FeSe}},\ }\href
  {https://doi.org/10.1103/PhysRevB.91.041112} {\bibfield  {journal} {\bibinfo
  {journal} {Phys. Rev. B}\ }\textbf {\bibinfo {volume} {91}},\ \bibinfo
  {pages} {041112} (\bibinfo {year} {2015})}\BibitemShut {NoStop}%
\end{thebibliography}%

\end{document}